# Generation of flat-top pulsed magnetic fields with feedback control approach

Yoshimitsu Kohama[a)] and Koichi Kindo

*The Institute for Solid State Physics (ISSP), The University of Tokyo, Kashiwa, 277-8581, Japan*

We describe the construction of a simple, compact, and cost-effective feedback system that produces flat-top field profiles in pulsed magnetic fields. This system is designed for use in conjunction with a typical capacitor-bank driven pulsed magnet, and was tested using a 60-T pulsed magnet. With the developed feedback controller, we have demonstrated flat-top magnetic fields as high as 60.64 T with an excellent field stability of ±0.005 T. The result indicates that the flat-top pulsed magnetic field produced features high field stability and an accessible field strength. These features make this system useful for improving the resolution of data with signal averaging.

Keywords: pulsed magnetic field, micro-second feedback controller, flat-top field profile

[a)]Author to whom correspondence should be addressed.
Electronic mail: ykohama@issp.u-tokyo.ac.jp.

**I. INTRODUCTION**

Pulsed magnets of various designs are in widespread use today [1]. Usually, the pulsed field profile is not a major concern for fast measurement techniques, such as magnetization and electrical transport measurements, whereas some specific measurements require the use of a particular field profile. The flat-top pulsed magnetic field (FTPMF), which has a constant field region around the maximum field value is a demanding field profile especially for nuclear magnetic resonance[2] (NMR) and specific heat measurements[2,3] in which a sample needs to be equilibrated in time intervals of less than the duration of the FTPMF. To date, many different approaches have been investigated [4-10] to increase the magnetic field strength, stability, and duration of FTPMFs.

The rectification of a large current driven by an AC flywheel generator is a well-known way to obtain FTPMFs. This technique can provide both high magnetic field and long duration for a FTPMF (see the second and third entries in Table 1). For instance, the National High Magnetic Field Laboratory (NHMFL) at Los Alamos provides FTPMFs of up to 60 T with durations of 100 ms [4]. In recent years, Wuhan National High Magnetic Field Center (WHMFC) also installed a flywheel generator that produces FTPMF of 50 T for 100 ms [5]. However, the rectification of an AC current introduces ripples in the FTPMF: Its amplitude is typically ±0.25 T at several hundred Hertz [5]. As the ripples can be regarded as high $dH/dt$ noise, some of the measurement techniques sensitive to this $dH/dt$ noise, such as the magnetization measurement, cannot provide high-quality data in rectifier-controlled pulsed magnetic fields.[6]

In contrast, the field profile of capacitor-bank-driven pulsed magnetic fields is perfectly smooth and its field strength can surpass 60 T. However, this type of pulsed field does not strictly have a flat region in the field profile; specific experiments that use such FTPMFs are difficult to perform. Even at the High Field Laboratory in Dresden (HLD), which generates the longest pulsed duration of 1.5 s for a capacitor-bank-driven pulsed magnetic field, the field strength changes by ±1 T over intervals of 70 ms [2]. To overcome this issue, a modification of the pulsed field profile has been investigated by dividing the capacitor bank modules into two groups: one driving the current for the pulsed magnet and the other suppressing the current around field maxima, thereby leveling the top of the field pulse [7]. This technique can make a highly stabilized FTPMF at 41 T (±0.09 T) for 6 ms.

In this work, we examined a feedback control strategy to obtain a stabilized FTPMF at high pulsed magnetic fields. Around the field maximum of the 60-T capacitor-driven pulsed fields, a homemade feedback controller regulates small magnetic fields of a 1.3-T mini-coil at intervals of 2.5 μs. Under feedback control, a stabilized FTPMF (60.64 ± 0.005 T) was obtained for 2.0 ms.

## II. DESCRIPTION OF THE FEEDBACK SYSTEM

### A. Coil Setup

Our experimental apparatus includes a mini-coil [see Fig. 1(a)] made of copper wire (ϕ1 mm) wound around a 12-mm-diameter stainless steel tube (0.3 mm thickness). The coil consists of a 37-mm-long main coil and two 20-mm-long oppositely-wound compensation coils, each composed of two layers and tightly wound solenoids. To prevent deformation of the coil caused by the electromagnetic force, the mini-coil is impregnated with epoxy resin (Stycast 1266) and is wrapped with Kapton tape. The resultant inner and outer diameters of the mini-coil are 11.4 and 17.5 mm, respectively. The coil has an inductance of 30 μH, and the resistances of the mini-coil are 130 and 20.8 mΩ at 300 and 77 K, respectively.

The mini-coil is inserted to the 60-T user coil [see Fig. 1(b)], which has a 18.0-mm inner bore diameter. The mini-coil is positioned on the field center of the user coil and rigidly clamped to it. As in Fig. 2, the user coil is driven with a 900-kJ (18 mF and 10 kV) capacitor bank and can generate magnetic fields up to 60.5 T with a pulsed duration of 36 ms.

### B. Electronics

To regulate the field profile within the short time scale of the pulsed fields, we constructed a fast feedback controller with a combination of metal-oxide-semiconductor field-effect-transistors (MOS-FET), lead acid batteries and a field-programmable gate array (FPGA) module. The electronic circuit inside the gray dotted lines in Fig. 2 is the circuit diagram of our feedback system. The current source is two 12-V lead-acid batteries (KUNG LONG; WP 50-12NE) connected in series. The current on the mini-coil circuit is controlled with the gate voltage in the four power-MOS-FETs (International Rectifier; IRFP 2907 PBF), where the MOS-FET works as a variable resistor with resistances ranging from ~3.6 mΩ to more than 10 MΩ enabling the current on the mini-coil circuit to be regulated. The FPGA module (National Instruments USB-7856R) controls the value of the gate voltage by comparing the field strength with the target field in a feedback loop (2.5 μs each). In a single feedback loop, this module completes the following four procedures; (1) collecting from the pickup coil the $dH/dt$ signal which is proportional to the time derivative of magnetic field, (2) converting the $dH/dt$ signal into a magnetic field, (3) calculating the appropriate gate voltage based on the feedback algorithm (discussed below), and (4) applying the gate voltage to the MOS-FETs. The immediate response of the FPGA module enables a microsecond time-scale feedback control of magnetic field.

**C. Feedback algorithm**

Our feedback algorithm is based on the well-known proportional-integral-derivative (PID) control method [11]. The value of the gate voltage ($V_{gate}$) is calculated using,

$$V_{gate}(t) = K_P \mathrm{e}(t) + K_I \int_{t_0}^{t} \mathrm{e}(t)dt + K_D \frac{d}{dt}\mathrm{e}(t), \qquad (1)$$

where the first, second, and third terms are the proportional, integral, and derivative terms, respectively. Each of these terms is dependent on the error value, $e$, which is the difference between the magnetic field at the present time and the target field ($H_{target}$), $e(t) = (H(t)-H_{target})$. $K_P$, $K_I$, and $K_D$ in Eq. (1) are called the proportional, integral, and derivative gains, respectively; $t$ and $t_0$ are the present and start times of the PID controller. Note that the magnetic field signal collected by the pick-up coil takes the form of $dH/dt$. Hence, the derivative term is obtained straightforwardly by multiplying $K_D$. To obtain the proportional and integral terms, the FPGA module performs the single- and double-integrations on each feedback loop, respectively.

**III. RESULTS AND DISCUSSIONS**

**A. Field profile generated by the mini-coil circuit**

Before performing any experiments with the present feedback system in pulsed magnetic fields, the performance of the mini-coil circuit at zero magnetic field needs to be checked. In Fig. 3, the dependence of the field profile on the fixed gate voltage is plotted. Because the drain-source resistance of the MOS-FET is reduced as $V_{gate}$ increases, the peak values of magnetic field and the current on the mini-coil circuit ($I_{mini}$) monotonically increase up to 1.31 T and 680 A, respectively, with $V_{gate}$ of 6 V. (The $I_{mini}$ data are not shown, but its profile is identical to the field data.) This result suggests that by suitably controlling the $V_{gate}$ from 4 to 6 V, we can obtain any waveform for magnetic field pulses up to ~1.3 T.

It is important to note that the time scales for increasing (up-sweep) and decreasing (down-sweep) the magnetic field are different. During up-sweeps, the magnetic field gradually increases within a 1–3 ms time scale (upper left inset of Fig. 3), followed by the application of gate voltage. In contrast, when the gate voltage is turned off, the magnetic field rapidly drops in ~100 μs (upper right inset of Fig. 3). This behavior can be roughly understood by assuming the mini-coil circuit to be a simple resistor-inductor (RL) circuit. As the time constant for an RL circuit is $\tau = L/R$, where $L$ and $R$ are the inductance and resistance of an electrical circuit, a reasonable $\tau$ of 1 ms is obtained for up-sweep setting $R = 30$ mΩ and $L = 30$ μH. To understanding down-sweep, stray capacitances and response times of MOS-FETs need to be taken into account, but the simple RL model can qualitatively account for the drop in $\tau$ as a consequence of an increase in $R$. This result indicates that the present feedback system has limits imposed on the field sweep rates of ~0.3–1 Tms$^{-1}$ (up-sweep) and ~10 Tms$^{-1}$ (down-sweep).

### B. Flat-top pulsed magnetic fields

With the fast feedback system explained in Section II, we tested the control of the pulsed field profile to obtain FTPMFs. Figure 4 exhibits the time dependences of $H$, $dH/dt$, $V_{gate}$, and $I_{mini}$. At point 1 in Fig. 4, the $V_{gate}$ of 5.5 V is applied and the subsequent increases in $H$, $dH/dt$ and $I_{mini}$ are observed. The 60-T field pulse starts at point 2, after which $H$ and $dH/dt$ rapidly increase. The $I_{mini}$ is also affected by the pulsed field generation due to the uncompensated portion of the induction voltage. At point 3, at which time the field reaches the target magnetic field, the FPGA starts to reduce $V_{gate}$ to produce the FTPMF. Following a reduction in $V_{gate}$, $dH/dt$ suddenly drops to zero, and $H$ exhibits a flat region. Here the flatness of the pulsed magnetic field can be clearly seen in the enlarged plot (Fig. 5). The field strength keeps almost constant value around the target field until point 4, at which field control stops. The operations of the field and mini-coil circuits are finished at points 5 and 6, respectively, and $H$, $dH/dt$, $V_{gate}$, and $I_{mini}$ all return to zero.

In Fig. 5(a–c), we plot $H$ (black) and $dH/dt$ data (red) around FTPMF with a comparison to those without a feedback control (blue dotted curves). The arrow in Fig. 5 indicates the timing at which the FPGA start/stop the control of the gate voltage. The field profile becomes very flat when the feedback control is turned on. Defining FTPMF as the field region having ±0.005 T stability, the time duration of the FTPMF is 2.0 ms with the field stability of ±0.008% at 60.64 T. Compared with the rectifier controlled pulsed field (50–60 ± 0.25 T)[4,5], the field stability and accessible field strength of the present system is remarkable.

Two points are noteworthy: (1) As frequently observed in a PID control method, an oscillation in the magnetic field is observed after ~300 μs of starting feedback control. We find that the degree of oscillation can be reduced/eliminated by an appropriate setting of the PID parameters ($K_P$, $K_I$ and $K_D$). Typically we choose $K_P = 0.25$ mVT$^{-1}$, $K_I$

= 4 Vs$^{-1}$T$^{-1}$, and $K_D$ = −1.5 mVsT$^{-1}$, and decreases in value can lead to a smaller oscillation. (2) From Fig. 5, the flat region of the magnetic field ends slightly before the feedback control stops. This is because the field sweep rate of the 60-T magnet rises above that of the mini-coil which has a slow field sweep rate (~0.3−1 Tms$^{-1}$) during field increases. This problem may be avoided using a high-voltage current source or a slow (small d$H$/d$t$) pulsed magnet. With the former, both the field strength and the field sweep-rate are expected to increase by applying a high-voltage to the mini-coil circuit, so that there is no doubt we have a FTPMF of longer duration. For example, a 770-V battery bank installed in the WHMFC might be suitable to operate the mini-coil [12]. With the latter, a combination of the present feedback system with a pulsed magnet having a long time duration, such as the 1.5-s long-pulse magnet at HLD[2], might be the most suitable to obtain a long-duration FTPMF. To check how the present feedback system works with small d$H$/d$t$ pulsed fields, we reduce the field strength generated by the 60-T pulsed magnet, and the results are presented in Fig. 6. Smooth FTPMFs within ±0.005 T instability are obtained with any pulsed field strength, where the PID parameters ($K_P$, $K_I$ and $K_D$) are kept constant. The time duration for the FTPMF is extended monotonically from 2.0 to 8.3 ms with a reduction in d$H$/d$t$ by a factor of ~1/6. Because the reduction in d$H$/d$t$ leads to a longer duration for FTPMF, we expect a long pulsed magnet with the present feedback system to be a powerful way to produce FTPMFs over long-time scales.

**IV. CONCLUSION AND OUTLOOK**

We have developed a FPGA-based feedback controller to obtain a flat region in a pulsed magnetic field. The present system is able to control the magnetic field over an interval of 2.5 μs, and hence leveling the pulsed field profile can be quickly achieved. This technique provides an improvement in field stability of a FTPMF by two orders of magnitude over previous methods using a rectifier. Furthermore, a FTPMF has been obtained at the unprecedented high magnetic field of 60.64 T. This system serves as a convenient apparatus for performing specific pulsed field measurements, such as specific heat [13] and NMR measurements [2], and also for other measurement techniques to improve the resolution of the data with signal averaging.

The present feedback system is driven only with two 12-V batteries. The number of batteries can be easily increased, which enhances both the field strength and field sweep rate of the mini-coil. Instead of MOS-FETs, an insulated-gate bipolar transistor (IGBT) can be used to deliver and control high currents and voltages in the order of ~kA and ~kV, respectively. Then, with a long magnet pulse, the pulse duration for the

FTPMF could be further extended.

**Acknowledgments**

The authors thank F. F. Balakirev and A. Kamoshida for useful comments on the FPGA module, and R. Otake and M. Akaki for help in winding the mini-coils. We are also grateful to X. Han, J. F. Wang and H. Kuroe for useful discussions. This work was supported by JSPS Grant-in-Aid for Scientific Research (C) Grant No. 15K05143.

Table 1. Summary of characteristics for flat-top pulsed magnetic fields generated worldwide.

| Facility | Maximum Field ($H_{max}$) | Stability ($H_s$) | Flat-top duration ($t_d$) | Reference |
|---|---|---|---|---|
| NHMFL | 60 T | — | 100 ms | [4] |
| WHMFC | 50 T | ± 0.25 T | 100 ms | [5] |
| HLD | 55.2 T | ± 1 T | 70 ms | [2] |
| WHMFC | 41 T | ± 0.09 T | 6 ms | [7] |
| ISSP | 60.64 T | ± 0.005 T | 2.0 ms | This work |

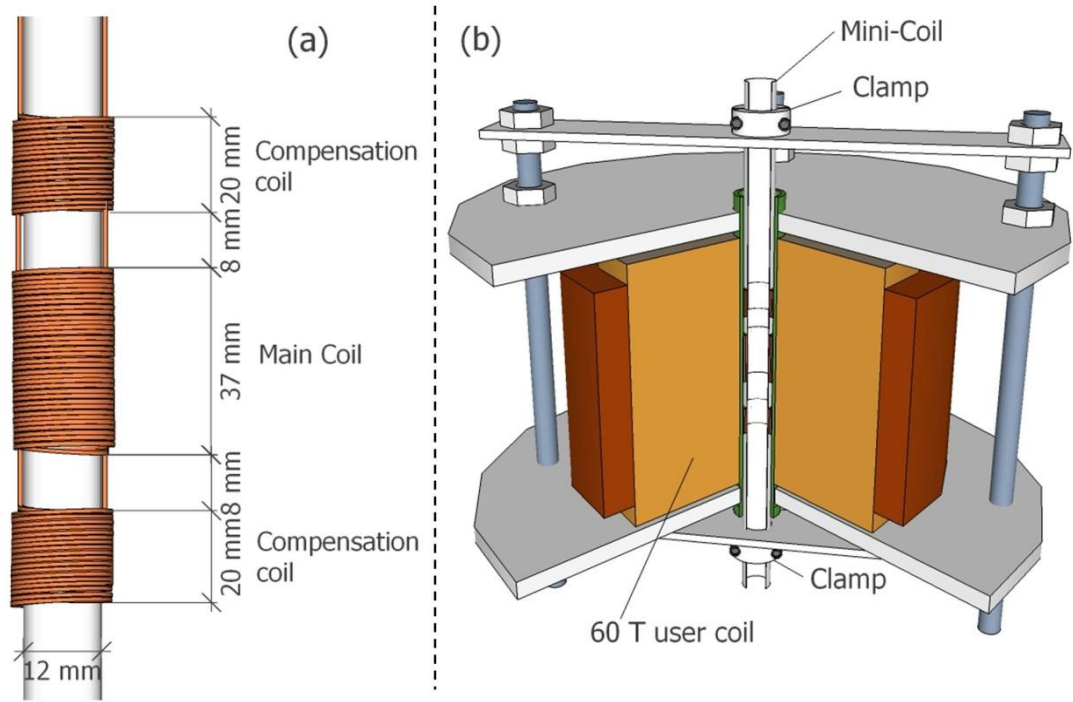

FIG. 1. (a) Sketch of the mini-coil. (b) Cutaway view of the mini-coil and 60-T user coil.

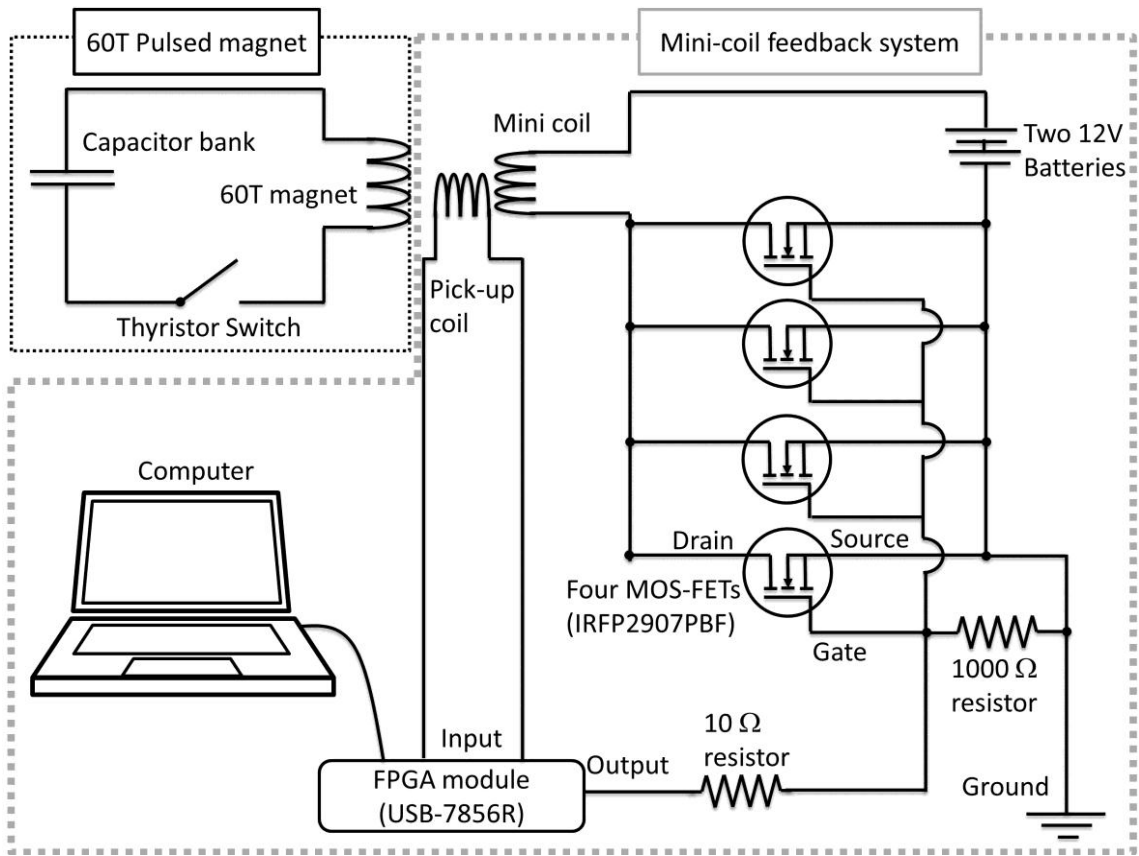

FIG. 2. Circuit diagram of our field feedback system developed in this work. The FPGA module contains analog input, analog output, and FPGA chip for an immediate onboard data processing. A personal computer is used to program and control the FPGA module. The 10- and 1-kOhm resistors constitute the gate and pull-down resistors, respectively.

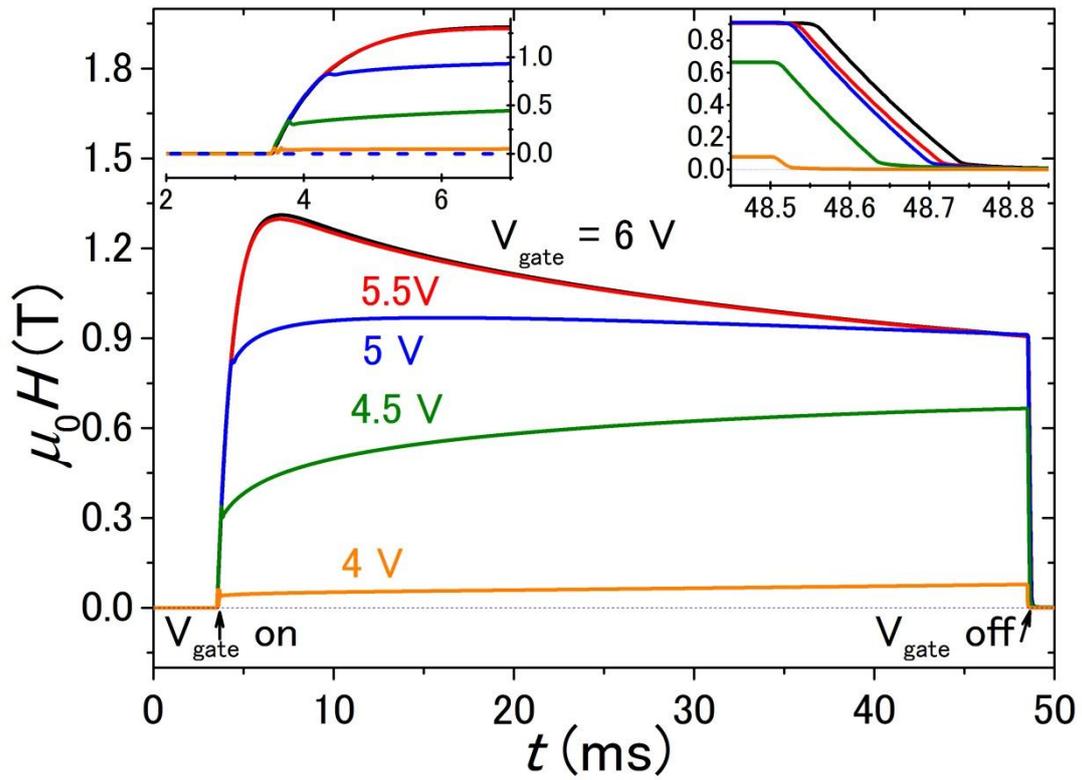

FIG. 3. Field profiles generated by the mini-coil circuit with constant gate voltages of 4 (orange), 4.5 (green), 5 (blue), 5.5 (red) and 6 V (black). Upper left and right insets show the enlarged parts at which V$_{gate}$ is turned on and off, respectively.

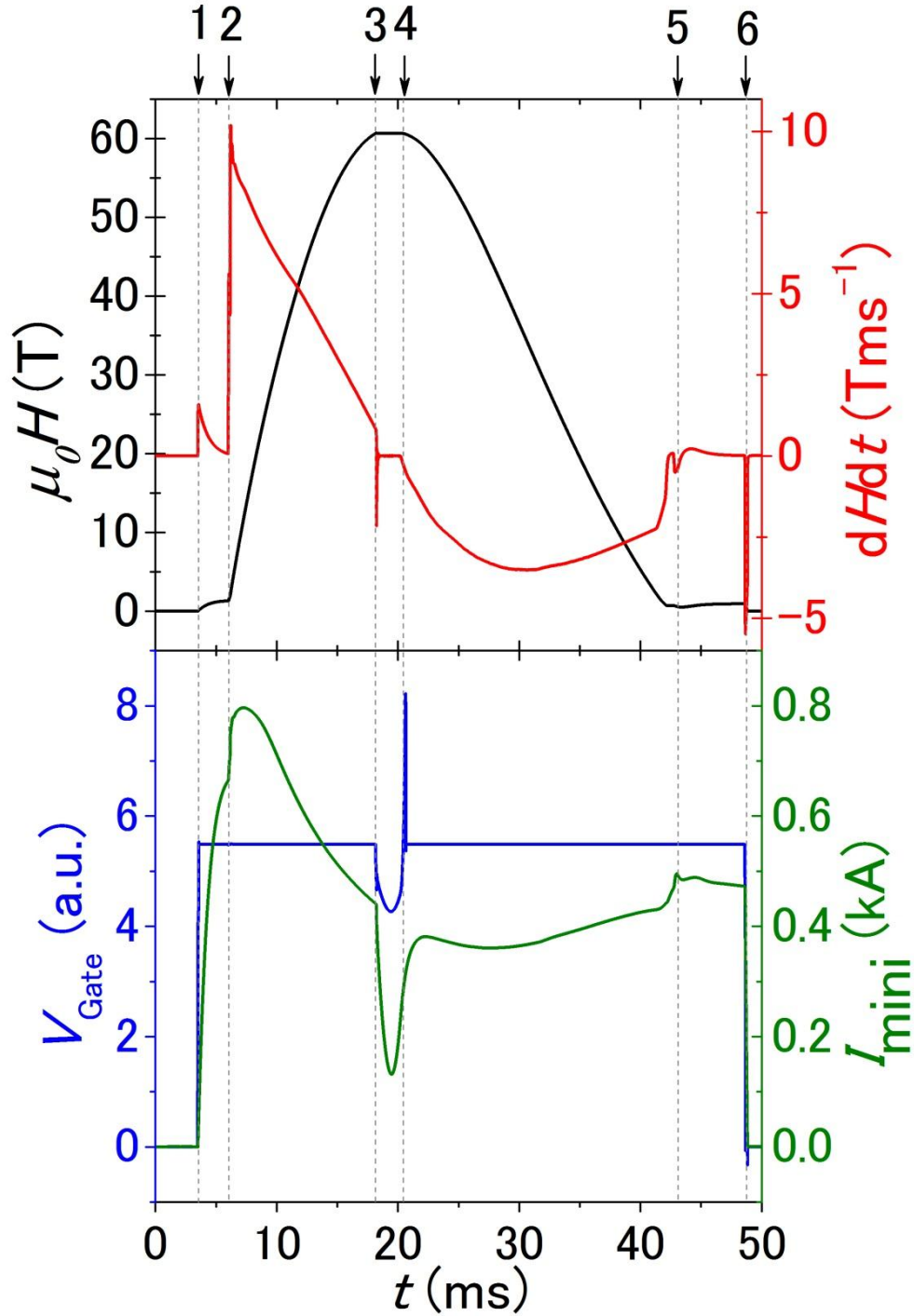

FIG. 4. Time dependences of the field profile (black), time derivative of the pulsed magnetic field (red), gate voltage (blue) and current in the mini-coil circuit (green). Arrows mark the sequence of events in the 60-T pulsed and mini-coil feedback circuits. (1) mini-coil circuit is turned on, (2) 60-T pulsed circuit is turned on, (3) FPGA starts the feedback control, (4) Feedback control is stopped, (5) 60-T pulsed circuit is turned off, and (6) mini-coil circuit is turned off.

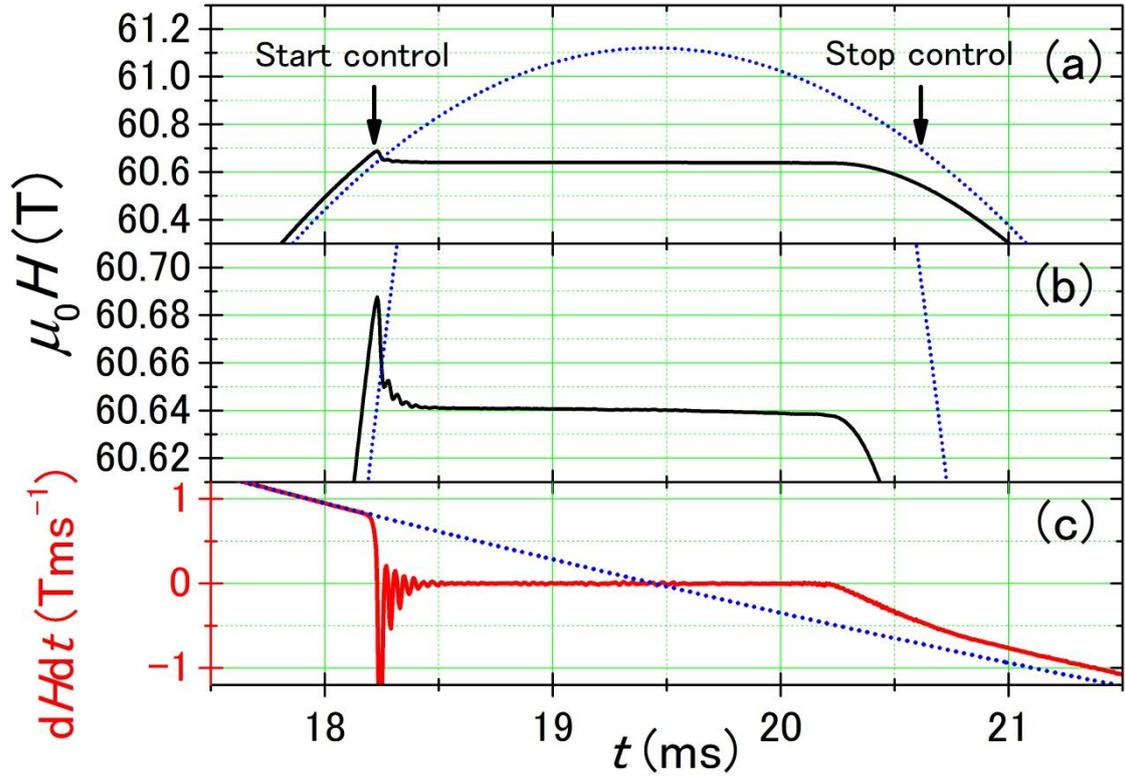

FIG. 5. Enlarged plots of the time dependences of $H$ and $dH/dt$ around the maximum of the magnet field pulse. The magnetic fields obtained with the feedback control are shown with black curve. The blue dotted curve is obtained with constant gate voltage of 5.5 V. (a) Enlarged plot from 60.3 to 61.3 T (±0.5 T) (b) Plot from 60.61 to 60.71 T (±0.05 T). (c) The $dH/dt$ signal with a feedback control is plotted as red curve.

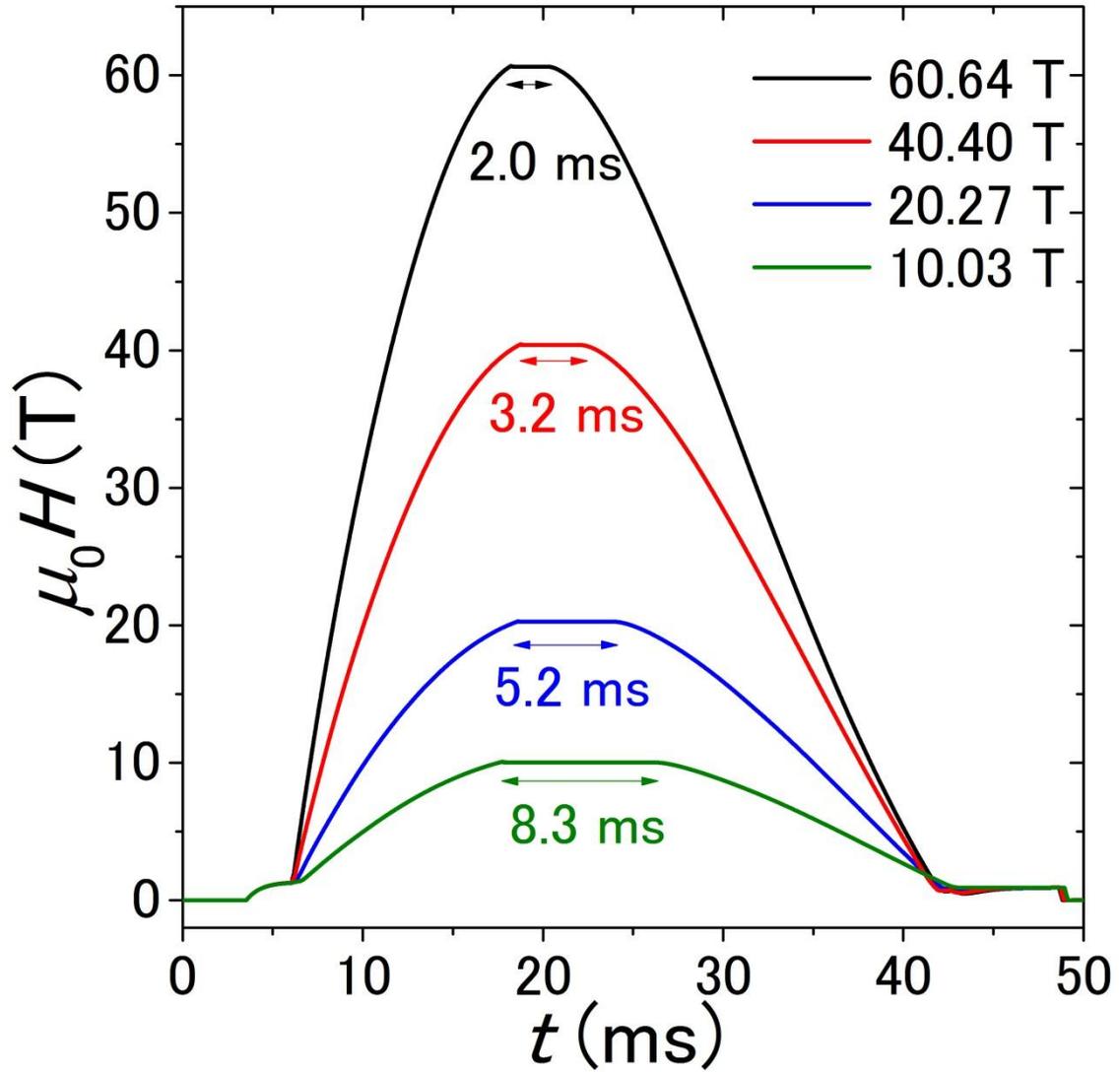

FIG. 6. Flat-top pulsed magnetic fields obtained with the developed feedback controller. The arrows indicate the duration of each flat-top magnetic field. The black, red, blue, and green curves are the flat-top magnetic fields at 60.64, 40.40, 20.27, and 10.03 T for 2.0, 3.2, 5.2, and 8.3 ms, respectively. For all flat-top magnetic fields, we use the same PID parameters. The duration of the flat-top pulsed magnetic fields is defined as the time interval when the magnetic field is stabilized within ±0.005 T.